\documentclass[aps,prl,twocolumn,groupedaddress,showpacs,floatfix]{revtex4-1}

\usepackage{graphicx}% Include figure files
\usepackage{bm}% bold math
\usepackage{amsmath}
\usepackage{float}
\usepackage{xspace}

\newcommand{\InO}{In$_x$O$_y$\xspace}
\newcommand{\AlAlO}{Al-Al$_2$O$_3$\xspace}
\newcommand{\AlO}{Al$_2$O$_3$\xspace}

\newcommand{\Hstar}{H$^*$\xspace}
\newcommand{\Hmax}{H$_{max}$\xspace}
\newcommand{\Rmax}{R$_{max}$\xspace}
\newcommand{\Tc}{T$_c$\xspace}
\newcommand{\Bc}{B$_c$\xspace}
\newcommand{\Hres}{H$_{90\%}$\xspace}
\newcommand{\E}[1]{x10$^{#1}$\xspace}

\begin{document}

\title{Onset of Nernst Effect Beyond the Coherence Critical Field of a Nano-Scale Granular Superconductor}

\author{S. Lerer}
\email[]{shaharl3@post.tau.ac.il}
\affiliation{Raymond and Beverly Sackler School of Physics and Astronomy, Tel-Aviv University, Tel Aviv, 69978, Israel}

\author{N. Bachar}
%\email[]{}
\affiliation{Raymond and Beverly Sackler School of Physics and Astronomy, Tel-Aviv University, Tel Aviv, 69978, Israel}

\author{G. Deutscher}
%\email[]{guyde@post.tau.ac.il}
\affiliation{Raymond and Beverly Sackler School of Physics and Astronomy, Tel-Aviv University, Tel Aviv, 69978, Israel}

\author{Y. Dagan}
%\email[]{yodagan@post.tau.ac.il}
\affiliation{Raymond and Beverly Sackler School of Physics and Astronomy, Tel-Aviv University, Tel Aviv, 69978, Israel}

\date{\today}

\begin{abstract}
We report measurements of the Nernst effect and of the magneto-resistance of granular aluminum films near the metal to insulator transition. These films show sharp transitions as a function of temperature and magnetic field. At low temperatures the Nernst signal displays a sharp peak at a field where more than 90\% of the normal state resistance has been restored, suggesting a transition involving entropy transport after superconducting coherence has been destroyed. At temperatures higher than the critical temperature the fluctuation paraconductivity scales with the Nernst signal, in agreement with a description in terms of fluctuations of the order parameter.
\end{abstract}

\maketitle
\section {Introduction}
The discovery of a strong Nernst voltage at fields thought to be higher than the accepted critical field value in several High \Tc cuprates\cite{Wang2006} has prompted a renewed interest in a phenomenon that had been studied in detail in conventional Type II superconductors\cite{Huebener1969}. In that case entropy is transported in the core of vortices moving under a temperature gradient, this motion resulting in a voltage transverse to the direction of the temperature gradient and of the applied field. Entropy transport vanishes as the upper critical is approached because, as the order parameter reduces, there is no more difference between the entropy in vortex cores and in the surrounding superconductor. This results in the disappearance of the Nernst signal at or before the upper critical field is reached.

%Figure 1
\begin{figure}[h!]
  \centering
  \includegraphics[width=0.81\linewidth]{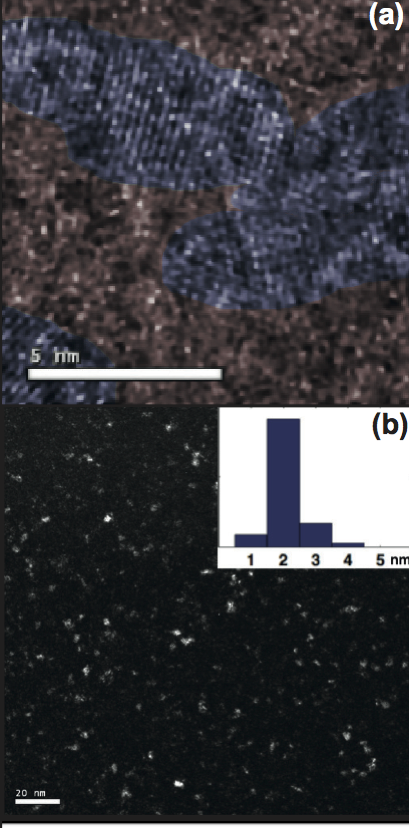}
  \caption{(a) Pseudo-Colored bright field TEM image of a typical sample. Red indicates a disordered region (amorphous \AlO) while blue indicates an ordered region (crystaline Al). (b) Dark field TEM image of a typical sample. Inset shows the distribution of the diameter of the grains (nm) in this image.}
  \label{fig:TEM}
\end{figure}
%Figure 1

%Figure 2
\begin{figure*}
  \centering
  \includegraphics[width=1\linewidth]{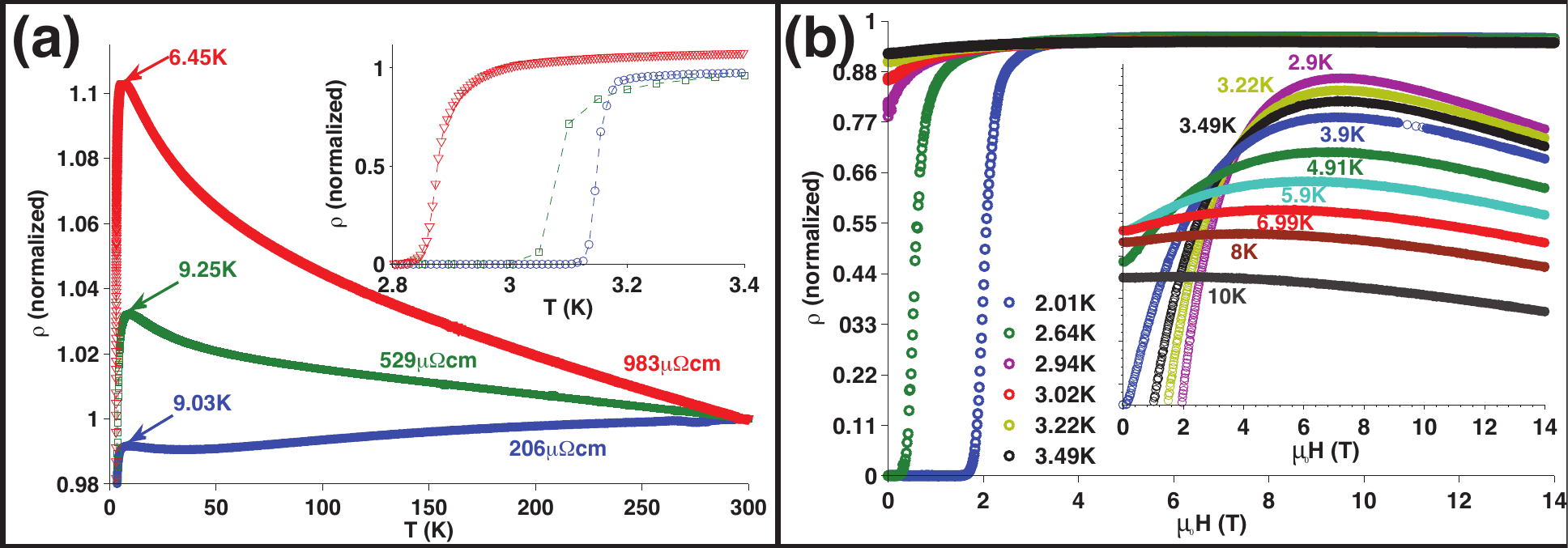}
  \caption{(a) Resistivities of the samples normalized to its value at 300~K. Inset shows a zoomed-in view on the resistive transition to the superconducting phase. (b) Normalized resistivity of sample C (983$~\mu\Omega$cm) as a function of the magnetic field. Inset focuses on the maximum in the resistivity.}
  \label{fig:Transport}
\end{figure*}
%Figure 2

%Figure 3
\begin{figure}
  %\centering
  \includegraphics[width=1\linewidth]{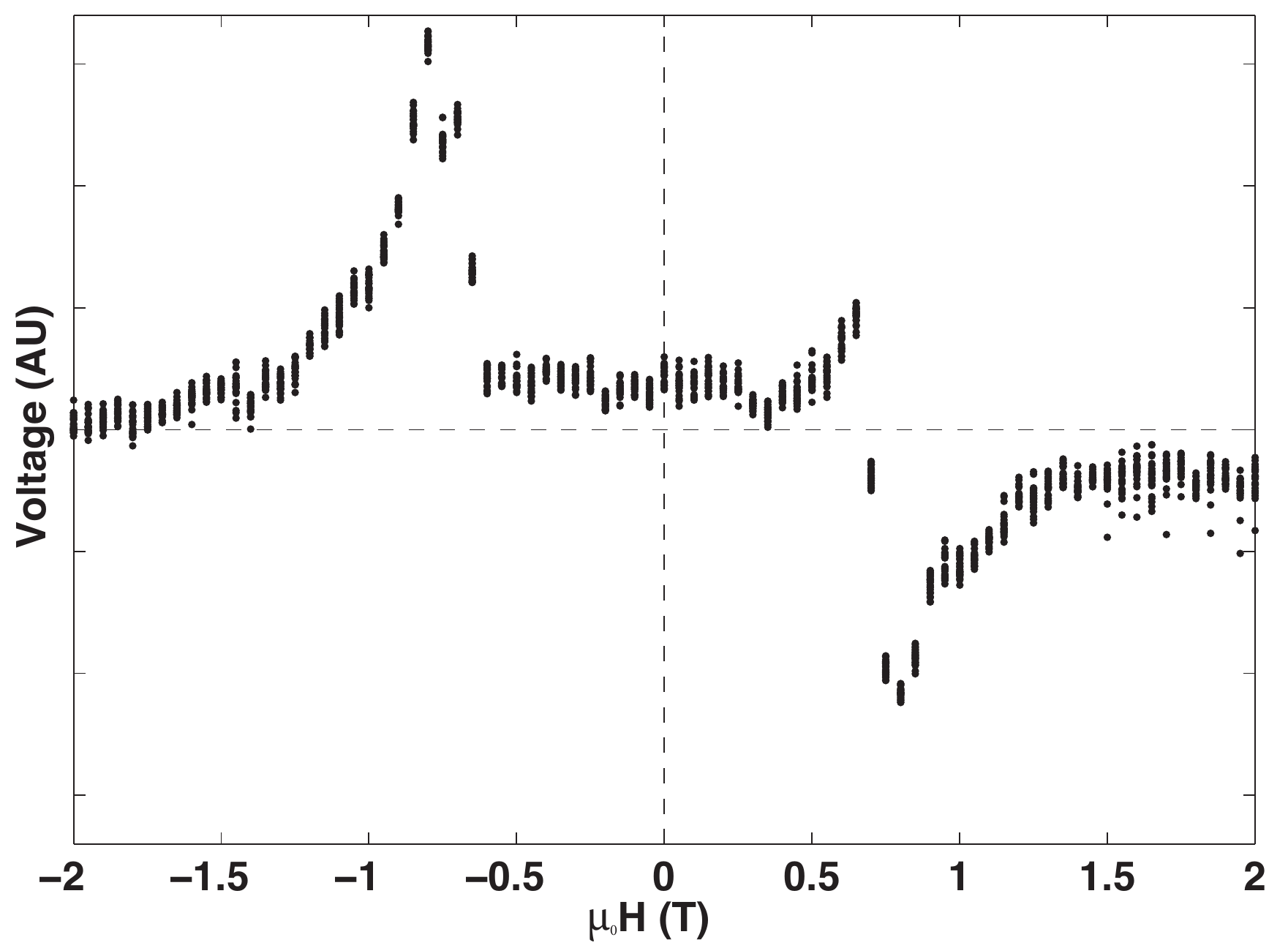}
  \caption{Raw Nernst voltage measurement for sample A at 2.12~K. For each field, the SC magnet was set to a 'persistent' mode and approximately 40 data points were measured at constant intervals in time. This data was then averaged (for each field value) and the symmetric part was removed. Afterwards, the data was normalized by the temperature gradient and the width of the sample.}
  \label{fig:RAW_N}
\end{figure}
%Figure 3

%Figure 4
\begin{figure}
  %\centering
  \includegraphics[width=1\linewidth]{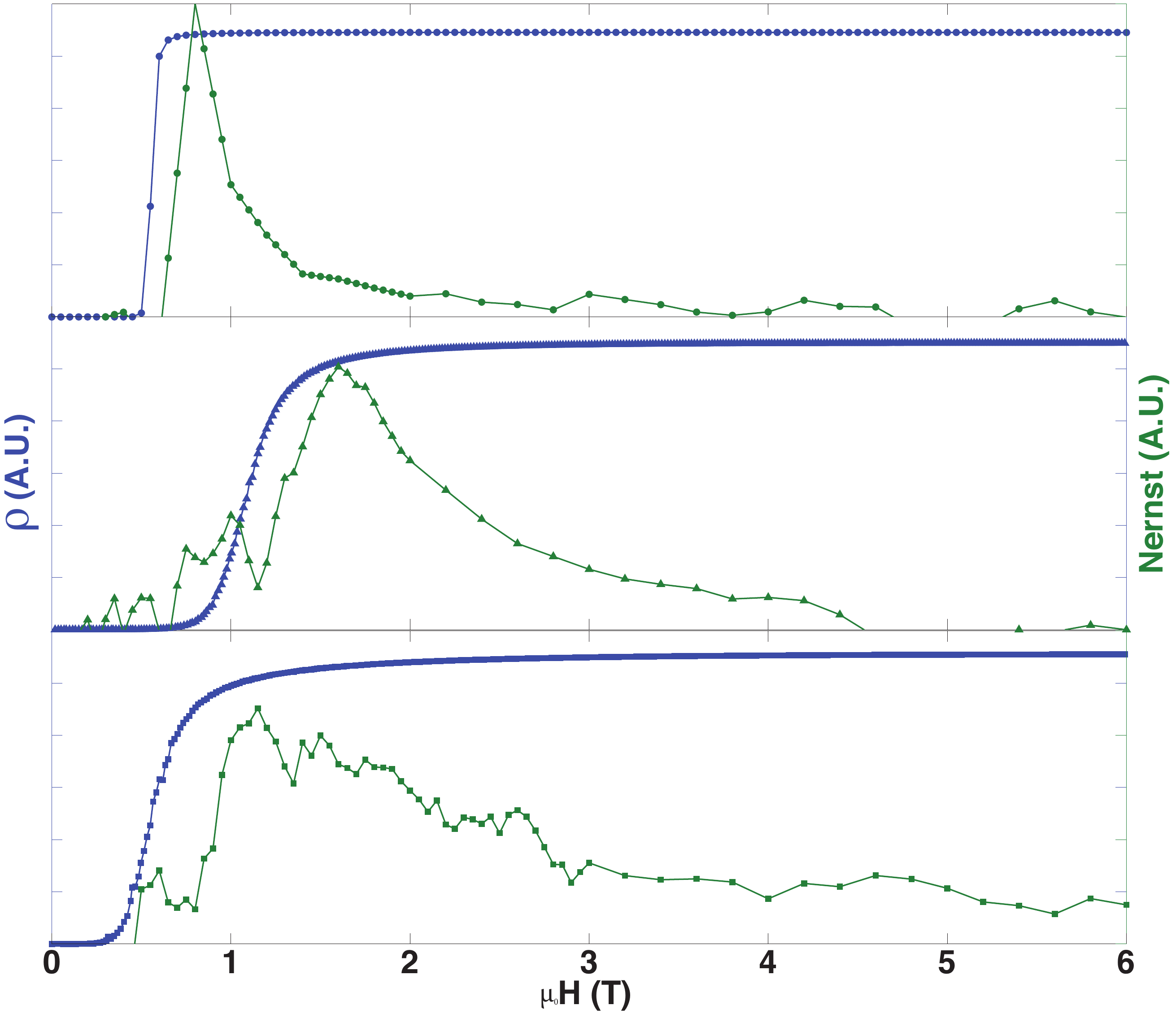}
  \caption{Behavior of the Nernst signal compared to the resistive transition for all three samples (A-C). The temperatures and thermal gradients are T$_A\sim$2.12~K, T$_B\sim$2.28~K, T$_C\sim$2.64~K and $\Delta$T$_A\sim$0.44~K/cm, $\Delta$T$_B\sim$1~K/cm, $\Delta$T$_C\sim$1.44~K/cm respectively. The Nernst signal is negligible in the superconducting phase and reaches its peak at a field where 90\% or more of its normal state resistance has been restored. It persists up to about 7T for sample C. }
  \label{fig:N_MR}
\end{figure}
%Figure 4

%Figure 5
\begin{figure*}
  %\centering
  \includegraphics[width=1\linewidth]{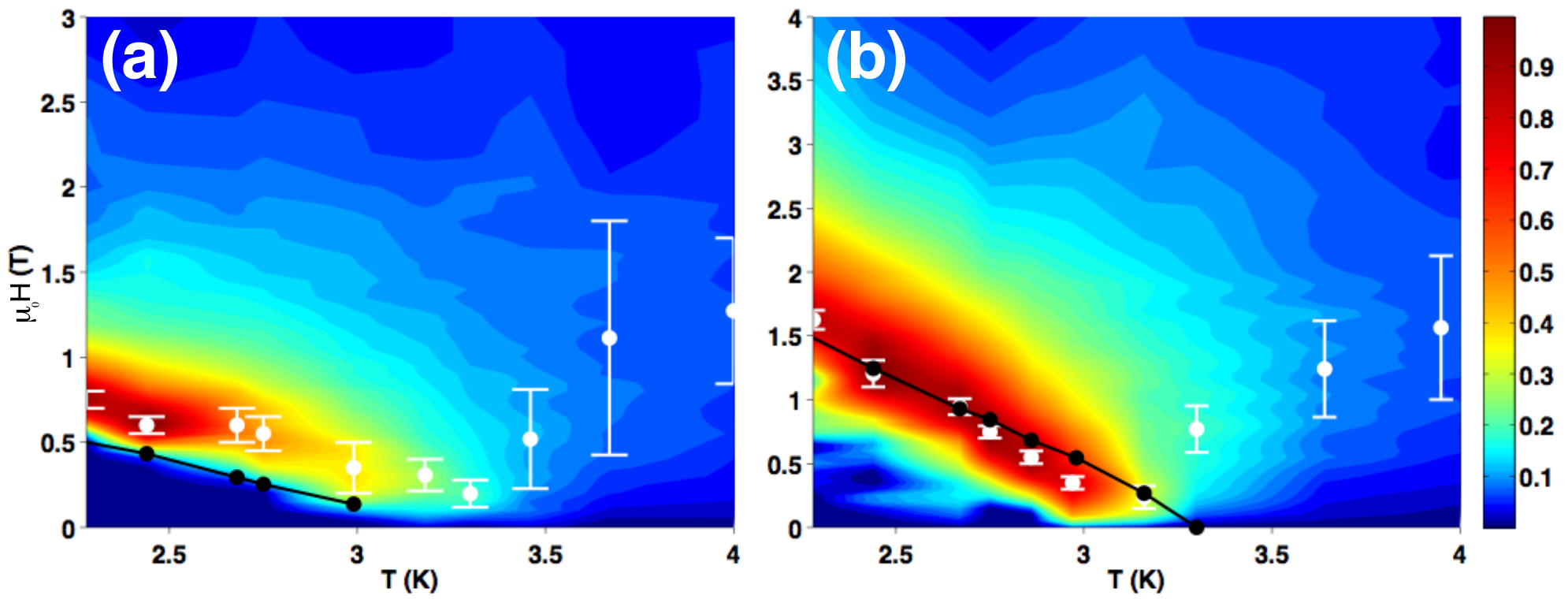}
  \caption{2D logarithmic heat-map of the Nernst signal measured for samples A (5a) and B (5b). Overlaid are the magnetic fields for the restoration of 90\% of the normal state's resistivity (\Hres) in black and the peak of the Nernst signal (\Hstar) in white.}
  \label{fig:NvsTvsH}
\end{figure*}
%Figure 5

%Figure 6
\begin{figure}
  %\centering
  \includegraphics[width=1\linewidth]{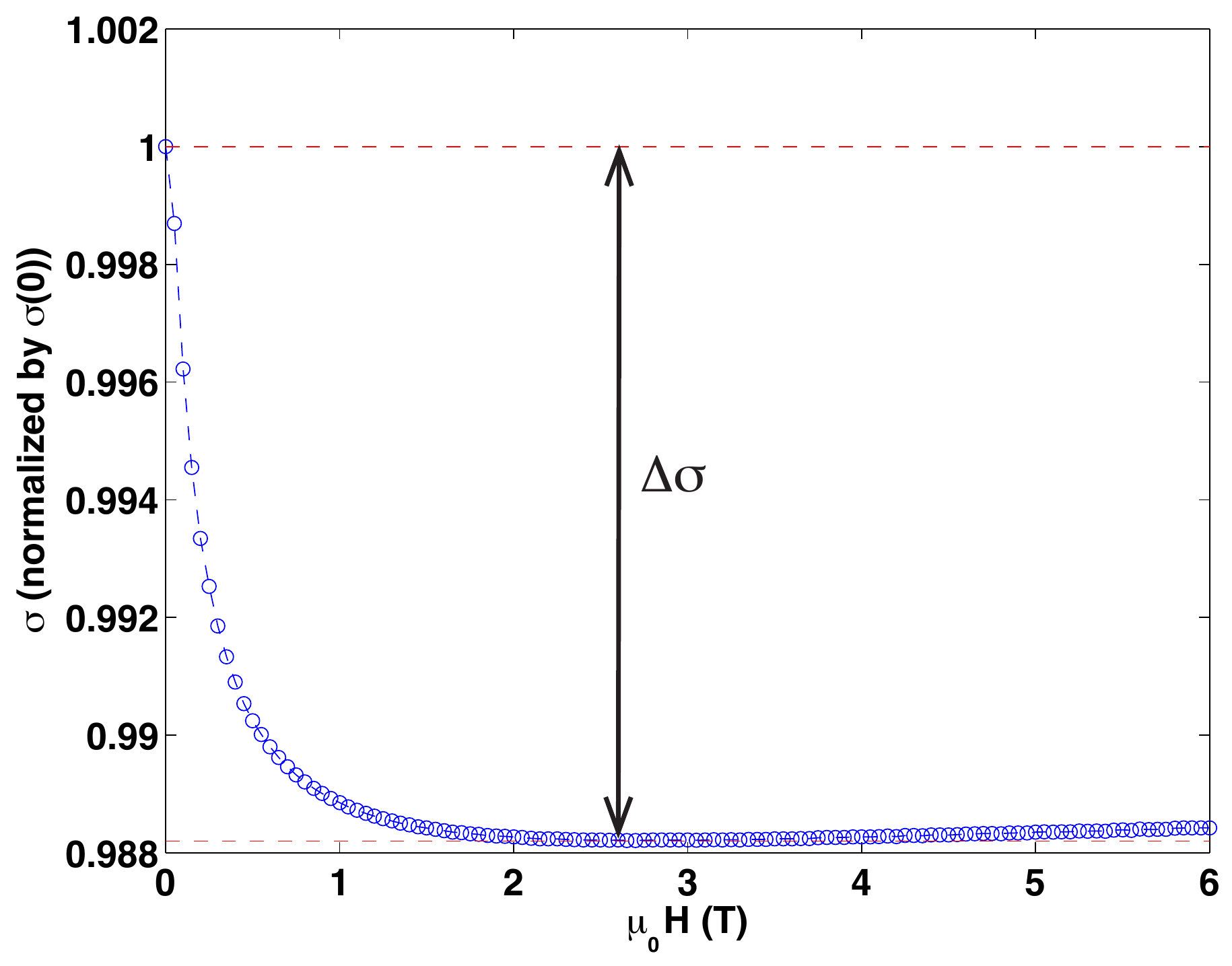}
  \caption{Normalized magnetoconductivity for sample A (206$~\mu\Omega$cm), at 3.46~K. The minimum was taken as the normal-state conductivity.}
  \label{fig:Delta_Rho}
\end{figure}
%Figure 6

%Figure 7
\begin{figure*}
  %\centering
  \includegraphics[width=0.9\linewidth]{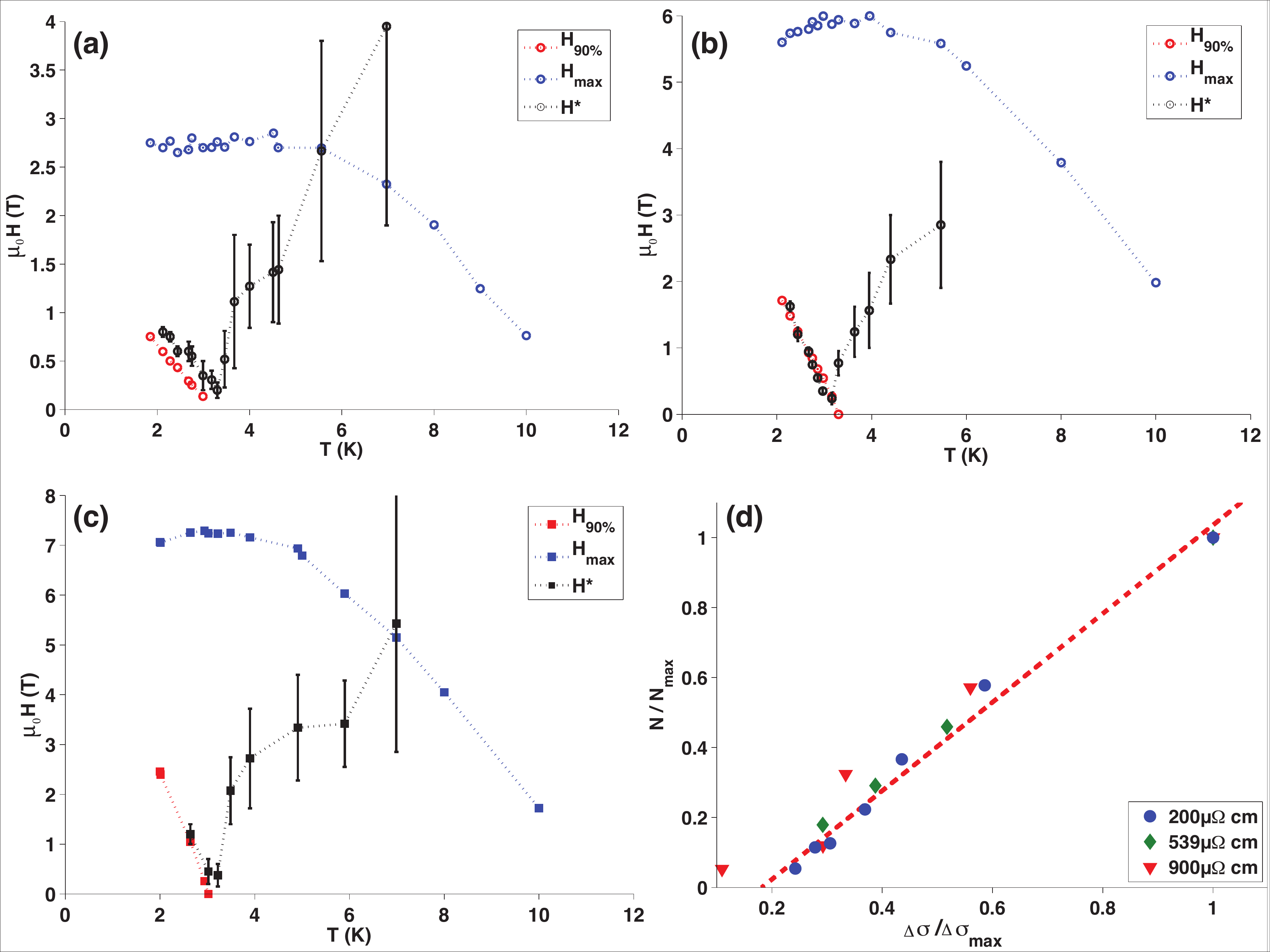}
  \caption{(a-c) Typical extracted magnetic fields for samples A-C. The fields are \Hres (the magnetic field for the restoration of 90\% of the normal state's resistivity), \Hmax (the fields where the magnetoresistance reaches a maximum and changes from positive to negative) and \Hstar (the peak of the Nernst signal). (d) Scaling of the added conductivity (magnetoconductivity) due to SC fluctuations and the maximal (peak) value of the Nernst signal for the same temperature, for all three samples (A-C). The data was normalized for the maximal value (added conductivity for each sample. Temperature range is between 3-5~K, where the Nernst signal is still significant compared to the background noise, and a peak is easily distinguishable. Also, the conductivity doesn't vary by more than 10\% of the value at $\sim$5~K. The dashed red line is a guide to the eye.}
  \label{fig:Phase}
\end{figure*}
%Figure 7

\par
The persistence of a strong Nernst signal at higher fields seen in the cuprate superconductors prompted the thought that some kinds of vortices subsist in these materials above their conventional upper critical field. But it was pointed out that in these materials a significant Nernst effect also exists in the normal state, which complicates the analysis of the data\cite{Taillefer2014}. To clarify the matter the Nernst effect was studied in highly disordered films where in the normal state the Nernst voltage is negligible. It was then discovered that in a disordered but otherwise conventional superconductor such as amorphous NbSi films the Nernst signal persists well above the critical temperature \cite{Behnia2009}. This persistence was explained in terms of Gaussian fluctuations of the order parameter\cite{Ussishkin2002,Michaeli2009}, and it was proposed that the origin of the Nernst signal seen at high fields and high temperatures in the cuprates was of a similar nature, namely that it did not imply the existence of vortices above the upper critical field.
On the other hand, it has been suggested \cite{Fisher1990} that a film close to a metal to insulating transition is prone to quantum fluctuations in the vicinity of a field tuned superconductor-to-insulator transition \cite{Hebard,YazdaniKapitulnik,goldman1998superconductor,Shahar} at zero temperature. In this scenario the language of vortices is indeed the proper one to be used to explain the persistence of the Nernst signal above the critical magnetic field \Bc. It has been suggested that such a description may be applicable also to the cuprates. \cite{steiner2005}
\par
The problem with the strongly disordered superconductors in which the Nernst signal has been so far studied is that they do not have a well defined critical field at which the resistance returns reasonably abruptly to its value in the normal state. It is therefore difficult to point out whether the Nernst signal does or does not persist in the normal state. This difficulty is particularly evident for \InO, where the transitions are extremely broad.
\par
In the experiment reported here we choose to measure the Nernst effect in granular Aluminum films, which have practically zero signal in the normal state enabling us to perform our measurements on top of a vanishingly small background. Growth conditions were optimized such that the resistive transition in zero field ($\frac{\Delta T_c}{T_c} \simeq 5\%$) and applied fields is sharp. It then becomes possible for us to compare the resistive and the Nernst transitions in a meaningful way. In addition, granular Al films have a very well characterized structure with a controlled narrow grain size distribution\cite{Deutscher1973}, which allows to discuss clearly the question of granularity that has been raised concerning other films suggested to be amorphous \cite{Ovadyahu1994}.

\par
We found that in the superconducting state upon increasing the magnetic field first electrical resistance appears and no Nernst signal is observed. The Nernst signal reaches its peak only when over 90\% of the normal state resistance has been recovered, and persists up to fields many times higher than the resistive critical field. This means that at first superconducting coherence is completely destroyed and only at higher fields entropy transfer becomes possible.

\par
\section {Sample preparation and experimental set-up}
Three granular aluminum (\AlAlO) samples were thermally evaporated at a rate of $\sim$6~\AA/sec, with different oxygen partial pressure varying from 2.4\E{-5} to 3\E{-5}~Torr, resulting in various resistivities. The thickness of the samples is 100~nm, except for the intermediate pressure sample, which has thickness of 50~nm.
\par
In Fig. \ref{fig:TEM}a (bright field) and \ref{fig:TEM}b (dark field) we show transmission electron microscopy images of a typical sample. The grain size is highly homogeneous and of the order of 2~nm (inset). Crystallite grains are embedded in an insulating amorphous oxide. This special combination of uniform sized grains gives rise to the unique conducting and superconducting properties that we report here.
\par
Electrical transport measurements were performed for all three samples. Nernst measurements were performed for all samples using the ‘diving board’ geometry\cite{Greene1997}, where the cold end of the sample was mechanically clamped onto a thermal bath, and the other end floating freely and heated with an SMT resistor.

\par
\section {Results and discussion}
In Fig. \ref{fig:Transport}a we show the resistance normalized with its value at 300K versus temperature for all samples: 206$~\mu\Omega$cm (Sample A) a metal-type, 529$~\mu\Omega$cm (B) and 983$~\mu\Omega$cm (C) an insulating-type. All of them are close to the metal-to-insulator transition exhibiting sharp superconducting transitions (see inset).

\par
In Fig. \ref{fig:Transport}b we present the normalized resistance versus magnetic field at various temperatures. The superconducting region at low fields is followed by a positive MR regime beyond which a negative MR region sets in (see inset).

\par
Fig. \ref{fig:RAW_N} shows an example of raw Nernst data (before the antisymmetrisation procedure). The sharp rise of the Nernst signal at a well defined field can be seen clearly.

\par
Fig. \ref{fig:N_MR} shows the field dependence of the resistivity and of the Nernst signal at low temperatures for the various samples. We define the field \Hres at which 90\% of the normal state resistance has been restored as the coherence critical field. Above \Hres macroscopic coherence over the entire sample is destroyed. Since most of the normal state sample resistance is due to the inter-grain tunneling barriers, and not to the normal state resistance of the Aluminum grains themselves, loss of coherence does not necessarily imply total destruction of superconductivity but just loss of inter-grain Josephson coupling. As shown in Fig. \ref{fig:N_MR} at low temperatures the Nernst voltage peaks at or beyond this field for all three samples. This peak is particularly sharp in sample A. We thus consider a picture where disconnected SC islands subsist beyond \Hres.

\par
The undetectable Nernst voltage below \Hres is in sharp contrast with the behavior of conventional superconductors. This behavior may be due to the fact that at lower fields vortices are mostly Josephson vortices located between superconducting islands. These vortices have no core. As shown previously both experimentally \cite{logvenov1994} and theoretically \cite{Mitin2010} moving Josephson vortices do not carry the usual Abrikosov  core entropy. Alternatively vortices may be strongly pinned up to \Hres, but we believe this to be unlikely, since the Nernst signal peaks at fields greater than \Hres.

\par
The emergence of a large and sharply peaked Nernst signal beyond \Hres implies that beyond this field vortices still exist, presumably Abrikosov vortices inside superconducting islands. Although they may be short lived their motion does carry entropy. In this respect we note that the resistive critical fields, of the order of 1~T at the lowest temperature where the Nernst effect could be measured in our set up, correspond to a coherence length of about 10~nm, several times the grain size. Hence, superconducting islands may consist of clusters of several grains.

\par
The magnetic field beyond which the Nernst effect cannot be detected with our set-up is considered as the field where superconductivity is quenched completely in the remaining islands. This field reaches up to 7~T. This rules out the previously measured resistive upper critical field of 3.6~T as being the Clogston-Chandrasekhar limit for granular films\cite{ChuiPaulilimit}.

\par Fig. \ref{fig:NvsTvsH} shows for samples A and B in color codes the behavior of the Nernst signal up to 4~K and 4~T. Additionally, it also shows the fields \Hstar (the peak of the Nernst signal) and \Hres.

\par 
Usually, when one attempts to extract the paraconductivity contributed by superconducting fluctuations above \Tc, the resistivity is measured twice - once with no external magnetic field, and the second pass is conducted with an applied external magnetic field that is sufficiently high that superconductivity is quenched. In our case, this modus operandi is problematic due to a negative MR in the normal state\cite{Nimi2013}. As one approaches the superconducting transition, a standard R(H) curve for a fixed temperature gives a puzzling picture. For H$\rightarrow$0 resistivity is reduced due to thermodynamic fluctuations. On the other hand, for H$\gg$H$_c$(T) (in our case, H=14~T), the resistivity also reduces. Inbetween these two ends, the resistivity reaches a maximum. Since this maximum marks the transition from the onset of SC at low fields to the normal state negative MR at high fields, we decided to take it to be the normal state resistivity (denoted by \Rmax for each temperature). Therefore, the SC contribution to paraconductivity, $\Delta \sigma$, is extracted for each temperature from $\sigma(0)-\sigma_{min}$, as shown in Fig. \ref{fig:Delta_Rho} for sample A (206$~\mu\Omega$cm), at 3.46~K. Paraconductivity obtained from the R(H) curves can be followed up to roughly 10K (three times \Tc). Fig. \ref{fig:Phase}d shows the normalized paraconductivity vs. the normalized amplitude of the Nernst signal for all samples from 3~K (above \Tc, within 10\% of the local maximum in resistivity) up to 5~K. The Nernst signal and the paraconductivity were normalized by their respective values at 3~K (denoted by N$_{max}$ and $\Delta \sigma_{max}$ respectively). The scaling of the two quantities confirms that they have a common origin. Such scaling is predicted for Gaussian fluctuations of the superconducting order parameter~\cite{Ussishkin2002, Michaeli2009}. However, the temperature dependence of the peak of the Nernst signal (Figs \ref{fig:NvsTvsH}, \ref{fig:Phase}) decays faster than the log(T) dependence expected for Gaussian fluctuations.

\par Fig \ref{fig:Phase}a-c  shows the temperature dependence of \Hres below \Tc, and of \Hstar below and above \Tc, for all three samples. In addition, it also shows the temperature dependence of the field \Hmax where R(H) reaches its maximum. This is also the field up to which the Nernst signal can be followed, up to about 6~K. It is apparent that, below \Tc, \Hstar and \Hres closely follow each other. As also observed previously in other disordered superconductors, the dependence of \Hstar above \Tc mirrors that below \Tc\cite{Kapitulnik1985,reviewBehnia}.

\par
\section {Summary}
The Nernst signal and the magnetoresistance above and below the resistive critical field \Hres and temperature \Tc were measured for granular aluminum samples. These samples exhibit sharp resistive transitions and vanishingly-small normal-state Nernst-signal. These properties enabled us to observe the onset of the Nernst signal at a field equal to or greater than that at which 90\% of the normal state resistance has been restored. The Nernst signal persists well above \Tc and \Hres. 
\par
The minimalist interpretation of the low temperature data does not exclude the more attractive one formulated in terms of a quantum critical point. In fact, the distinction may be of a semantic nature. In any case, the Nernst measurements presented here, together with the resistive measurements, suggest that at low temperatures a transition takes place at a field \Hstar where the Nernst signal shows a sharp peak and 90\% or more of the normal state resistance is restored. Only vortices that do not carry entropy are present at H$<$\Hstar, either because they have no core or because they are localized. Mobile vortices carrying entropy are present at H$>$\Hstar up to fields several times higher than \Hstar. At temperatures higher than the critical temperature the fluctuation conductivity scales with the Nernst signal, in agreement with a description in terms of thermodynamic fluctuations of the order parameter.\\
\begin{acknowledgments}
We would like to thank K. Michaeli and R.P. Huebener for many insightful discussions.
This research was supported by the Israel Science Foundation under grants 569/13 and 1421/08 GD acknowledges support from EOARD Award No. FA8655-10-1-
3011.
\end{acknowledgments}
\bibliographystyle{apsrev}
\bibliography{Nernst_on_AlAl2O3_Final}
\end{document}